# Dynamic synchronization of a time-evolving optical network of chaotic oscillators

Adam B. Cohen,[1,2] Bhargava Ravoori,[1,2] Francesco Sorrentino,[1,3] Thomas E. Murphy,[1,4] Edward Ott,[1,2,4] and Rajarshi Roy[1,2,5]

[1] *Institute for Research in Electronics and Applied Physics, University of Maryland, College Park, Maryland 20742, USA*

[2] *Department of Physics, University of Maryland, College Park, Maryland 20742, USA*

[3] *Università degli Studi di Napoli Parthenope, 80143 Napoli, Italy*

[4] *Department of Electrical and Computer Engineering, University of Maryland, College Park, Maryland 20742, USA*

[5] *Institute for Physical Science and Technology, University of Maryland, College Park, Maryland 20742, USA*

We present and experimentally demonstrate a technique for achieving and maintaining a global state of identical synchrony of an arbitrary network of chaotic oscillators even when the coupling strengths are unknown and time-varying. At each node an adaptive synchronization algorithm dynamically estimates the current strength of the net coupling signal to that node. We experimentally demonstrate this scheme in a network of three bidirectionally coupled chaotic optoelectronic feedback loops and we present numerical simulations showing its application in larger networks. The stability of the synchronous state for arbitrary coupling topologies is analyzed via a master stability function approach.

**In spite of the extreme sensitivity of chaotic orbits to small perturbations, it is known that, when properly coupled together, identical chaotic systems can synchronize[1]. In the synchronized state, each of the component systems follows the same chaotic orbit in lockstep. This somewhat surprising phenomenon is a basis for several proposed applications, including the use of chaotic signals in communication and sensing[2]. Thus, establishing and maintaining synchrony on a network of nonlinear oscillators is an important goal. In many of the proposed applications it is assumed that the network couplings are *a priori* known, and this assumed knowledge is used in designing the synchronization strategy. However, in practice such knowledge may be unavailable or imperfect, or the couplings may depend on unobservable external time-dependent parameters. In this paper, we address such situations via an adaptive strategy that uses the chaos synchronization phenomenon to learn and track such *a priori* unknown couplings even as a particular chaos synchronization application is simultaneously being carried out.**

---

## I. INTRODUCTION

While there has been substantial theoretical and numerical simulation work on adaptive techniques for chaos synchronization, there has been little work on the experimental realization[3] or quantitative consideration of their stability[4]. In Ref. 5, an adaptive technique was devised to achieve synchronization of chaos through a gradient descent strategy. In Refs. 3, 6, and 7, techniques based on appropriately defined Lyapunov functions have been proposed. In Refs. 8 and 9, adaptive strategies have been investigated to synchronize dynamical complex networks. In Refs. 10 and 11, we introduced an adaptive strategy that maintains synchronization on a time-evolving network of chaotic oscillators and reported numerical simulations of this strategy on a network of low-dimensional systems. In Ref. 12, we studied the stability of this adaptive strategy using a master stability function approach[13,14]. In Ref. 15, we presented an experimental demonstration of the maintenance of synchrony on a pair of unidirectionally coupled optoelectronic chaotic feedback loops. Here, we report an experimental study of adaptive synchronization on a network of high-dimensional chaotic units. In particular, we consider time-delayed chaotic optoelectronic oscillators, and we illustrate the applicability of an adaptive algorithm for sensing and simultaneously tracking changes in the coupling strengths of the network while maintaining synchrony. We present experimental measurements from a three node network and show numerical simulations from larger networks.

In attempting to understand the emergence of synchrony in ensembles of nonlinear oscillators, many models have been introduced to describe the interaction of individual ensemble members (network 'nodes') with other members of the ensemble. In our formulation below, our model can potentially include such effects as propagation time-delays of couplings, temporal variations in coupling strength, changes in network topology, and high-dimensional chaotic dynamics. Thus the method described here applies to a wide range of networks[16–19].

## II. ADAPTIVE SYNCHRONIZATION TECHNIQUE

In order to establish a theoretical framework for the adaptive synchronization method, we first consider a network of $N$ coupled identical discrete time systems that evolve according to

$$\mathbf{x}_i(n+1) = \mathbf{F}(\mathbf{x}_i(n)) + \mathbf{v}\frac{\alpha_0}{k_i} r_i(n), \tag{1}$$

where $\mathbf{x}_i(n)$ represents the state-vector of node $i$ at time $n$. The quantity $r_i(n)$ is a scalar signal representing a cumulative coupling received from other nodes in the network and is assumed to be of the form $r_i(n) = \sum_{j=1}^{N} A_{ij} H(\mathbf{x}_j(n))$. The vector function $\mathbf{F}(\mathbf{x})$ and scalar function $H(\mathbf{x})$ describe the internal and coupling dynamics respectively. $\mathbf{A} = \{A_{ij}\}$ is an $N \times N$ weighted adjacency matrix, $\alpha_0$ is an overall coupling strength, $\mathbf{v}$ is a vector that describes how the received coupling signal $r_i(n)$ is incorporated, and $k_i = \sum_{j=1}^{N} A_{ij}$ describes the net coupling into the $i$-th node. (Note that in our scheme, $k_i$ is always greater than zero so that division by $k_i$ in (1) is well-defined.) The adjacency matrix $\mathbf{A}$ specifies the topology of the network along with the associated coupling strengths of the network links and will play a key role in determining whether synchronization occurs. Eqs. (1) admit a globally synchronized solution, $\mathbf{x}_1(n) = \mathbf{x}_2(n) = \ldots = \mathbf{x}_N(n) \equiv \mathbf{x}_s(n)$, satisfying $\mathbf{x}_s(n+1) = \mathbf{F}(\mathbf{x}_s(n)) + \mathbf{v}\alpha_0 H(\mathbf{x}_s(n))$. Note that in the design of this system it has been assumed that each node $i$ processes its received signal $r_i$ with knowledge of the net coupling $k_i$ (see Eqs. (1)). But what can be done in applications in which the $k_i$ are *a priori* unknown?

Our synchronization algorithm is designed to achieve and maintain synchrony in situations in which value of $k_i = \sum_j A_{ij}$ is unknown and time-varying, e.g., due to unknown temporally varying environmental conditions affecting the $A_{ij} = A_{ij}(n)$. Thus we now write $k_i = k_i(n)$. Our goal will be realized by adopting a local real-time adaptive strategy at each node that weights the received signal based on an estimate of $k_i$ determined only from the known received signal $r_i(n)$. Thus we re-express Eqs. (1) as,

$$\mathbf{x}_i(n+1) = \mathbf{F}(\mathbf{x}_i(n)) + \mathbf{v}\beta_i(n)r_i(n), \tag{2}$$

with the goal of setting $\beta_i(n) = \alpha_0/k_i(n)$. We follow Refs. 10 and 11 and seek to minimize a time-averaged measure of the synchronization error at each node, where the measure is given by,

$$\psi_i = \langle[\beta_i r_i - \alpha_0 H(\mathbf{x}_i)]^2\rangle_{z_0}, \tag{3}$$

where $\langle G\rangle_{z_0} = (1-z_0)\sum_{m=0}^{\infty} z_0^m G(n-m)$ and $z_0$ is a smoothing factor that determines the temporal extent over which the averaging is performed. The time-window over which this exponentially-weighted moving averaging is performed is $(1-z_0)^{-1}$ samples. The minimum value for $\psi_i$ is 0, and this is attained only when $\beta_i r_i = \alpha_0 H(\mathbf{x}_i)$. By minimizing $\psi_i$ with respect to $\beta_i$ (i.e., $\partial\psi_i/\partial\beta_i = 0$), we obtain a real-time estimate of the weight factor:

$$\beta_i(n) = \alpha_0 \frac{\langle r_i H(\mathbf{x}_i)\rangle_{z_0}}{\langle r_i^2\rangle_{z_0}} = \alpha_0 \frac{p_i(n)}{q_i(n)}, \tag{4}$$

where we assume the channel variations $A_{ij}(n)$ are slowly-varying relative to the smoothing window[10]. From our definition of the above time average, we have that the numerator and denominator satisfy the following iterative equations:

$$\begin{aligned} p_i(n) &= z_0 p_i(n-1) + (1-z_0) r_i(n) H(\mathbf{x}_i(n)), \\ q_i(n) &= z_0 q_i(n-1) + (1-z_0) r_i^2(n). \end{aligned} \tag{5}$$

Upon synchrony, $p_i(n) \to k_i\langle H(\mathbf{x}_i)^2\rangle_{z_0}$ and $q_i(n) \to k_i^2\langle H(\mathbf{x}_i)^2\rangle_{z_0}$, for which the ratio in Eq. (4) attains the desired value.

While the results are described here for the case of a discrete time dynamical system, we note that this formalism can also be applied to continuous time systems[10].

## III. EXPERIMENTAL SETUP

In this paper, we consider an experimental network of three coupled optoelectronic oscillators[15,20–23] as depicted in Fig. 1a. The nodes are connected to one another over bidirectional fiber optic channels. The coupling strengths $A_{ij}$ are controlled by applying a voltage to an electronically variable optical attenuator and in this work the couplings were designed to be symmetric ($A_{ij} = A_{ji}$). Although these systems are traditionally described using delay differential equations, they can be cast into the framework of Eq. (1) by discretizing the time axis and constructing a state vector $\mathbf{x}_i$ comprising one time-delay worth of sample points[24]. Appendix A explains how the continuous time delay differential equation describing the optoelectronic feedback loop can be converted into discrete time equations that can be readily implemented using digital signal processing hardware. Appendix B explains how the resulting discrete time equations and adaptive synchronization method can be cast in the general form of Eqs. (2)-(5). At each node (Fig. 1b), a laser diode provides a constant input optical power to a Mach-Zehnder electro-optic intensity modulator. The optical power transmitted by the modulator is proportional to $\cos^2(u_i - \pi/4)$, where $u_i(n)$ is the normalized voltage applied to the modulator. The modulated signal is split to act as an internal feedback signal and broadcast as a coupling signal to the neighboring nodes. An optical circulator is used to both transmit the outgoing signal and receive a summed input signal from the neighboring nodes. Each node receives an aggregate signal $r_i(n)$ from its neighbors and does not have independent access to each of its neighbor's transmitted signals. A real-time digital signal processor in each node is used to dynamically rescale the received signal according to the algorithm described in Sec. II. A typical time-trace of an isolated node is plotted in Fig. 1c, which exhibits robust high-dimensional chaos[21,22].

## IV. MAINTENANCE OF SYNCHRONY

To demonstrate the effectiveness of the adaptive synchronization method, we experimentally investigated the case when one of the coupling coefficients in the network ($A_{12}$) changes abruptly. For the data plotted in Fig. 2a, the adaptive method was not employed. In this case, the network was initially adjusted to achieve synchronization, with $A_{13} = 0, A_{23} = 1.5$ and $A_{12} = 2.8$. At $t = 0$, $A_{12}$ is abruptly changed to 1.3, and synchrony is broken, as evidenced by a large synchronization error, measured as the average of the absolute pairwise differences $(|u_1 - u_2| + |u_2 - u_3| + |u_3 - u_1|)/3$. This is expected, since the coupled equations that describe the dynamics of the network without adaptation no longer admit a synchronous solution. In Fig. 2b, we explore the same situation, with the adaptive al-gorithm enabled such that the scale factors $1/\overline{k}_i$ are estimated in real-time. After a short adjustment time following sudden change of $A_{12}$ at $t = 0$, we see that the synchronous state is rapidly restored.

Moreover, by monitoring the three tracking signals $\overline{k}_1$, $\overline{k}_2$, and $\overline{k}_3$, we are able to determine ('learn') estimates of the individual elements of the symmetric adjacency matrix $\mathbf{A}$ by solving the three linear equations $\overline{k}_i = \sum_j \overline{A}_{ij}$ to obtain estimates, $\overline{A}_{12}$, $\overline{A}_{23}$, and $\overline{A}_{31}$, of the three unknowns $A_{12}$, $A_{23}$, and $A_{31}$. In the top trace of Fig. 2b, we display $\overline{A}_{12}$ which is in good agreement with the actual coupling strength $A_{12}$ used in the experiment (top trace of Fig. 2a). In Fig. 3, we consider a case where $A_{12}$, $A_{23}$, and $A_{31}$ are all nonzero and both $A_{12}$ and $A_{31}$ are simultaneously time-varying. Again we see that the algorithm maintains synchrony throughout this process and independently tracks disturbances in each channel. The dashed lines show the actual channel variations, and the solid curves represent the network's response. (Note that, to make the dashed lines visible, the solid lines have been artificially shifted upward by 0.05.)

The ability to obtain estimates of the coupling matrix $\overline{A}_{ij}$, may be interesting in that variations of the $\overline{A}_{ij}$ may be induced by environmental changes, and estimating temporal changes in the $\overline{A}_{ij}$ may thus be used as a means of sensing these environmental changes. We note, however, that in larger networks, the number of edges could exceed the number of network nodes. In this case, although our adaptive strategy still maintains synchrony through adjustment and tracking of the values of $\overline{k}_i$, knowledge of only the $\overline{k}_i$ is no longer sufficient to allow one to deduce estimates of the coupling matrix $\overline{A}_{ij}$. However, the adaptive strategy could still prove useful for sensing localize perturbations, especially in sparsely connected networks or when perturbations are not occurring simultaneously. We also note that a closely related scheme (Ref. 11) potentially allows estimates of all the $\overline{A}_{ij}$ in large networks. The work we have presented here provides the first experiments highlighting the ability to track and temporally localize specific disturbances in a network of synchronized chaotic oscillators. The experiment can be considered a proof-of-principle test for the application of coupled dynamical systems as a sensor network. In this prototype system, the nodes act collectively to learn about specific changes in their environment. The control signals $\overline{k}_i$, which maintain synchrony in the network, also contain practical information used to sense what occurs between the nodes. Depending on the setting, the couplings could be arranged in free-space rather than with fiber optic cables or could use transmitter and receiver antennas.

## V. STABILITY RANGE OF SYNCHRONOUS SOLUTION

In the literature on synchronization of chaotic systems, the stability of a synchronous solution for a network of oscillators can be evaluated using a low-dimensional equation that depends on the eigenvalues of the Laplacian matrix associated with the network topology[13,14]. We show here that this approach can be extended to the present case. However, in this case the adaptive rule becomes part of the dynamical equations of the system and must therefore be included when calculating the master stability function.

To experimentally investigate the range of stability, we considered the same adaptively controlled three-node network as in Fig. 2b (i.e., $z_0 = 0.99$, $A_{31} = 0$, $A_{23} = 1.5$), but we varied the third coupling parameter $A_{12}$ between 0 and 4. At each point, we measured the average pairwise synchronization error,

$$\sigma = \frac{1}{N(N-1)} \sum_{i,j,i \neq j} \sqrt{\frac{\langle (u_i - u_j)^2 \rangle}{\langle u_i^2 \rangle + \langle u_j^2 \rangle}}, \tag{6}$$

which is 0 for an identically synchronized network and 1 for a network of uncorrelated nodes. In Fig. 4a, the points are the experimentally obtained synchronization errors for $z_0 = 0.99$ and the corresponding solid curves are from numerical simulations. As indicated particularly by the simulations, there are upper and lower bounds for $A_{12}$ for which stability is achieved. The range of $A_{12}$ in which we observe synchronization can be explained by analyzing the stability of the synchronous solution using the master stability function technique[13,14,25]. We emphasize that in doing this it is essential to include the dynamics of the adaptive strategy[12]; e.g., the smoothing parameter $z_0$ has a substantial effect in determining stability. The stability is related to the $(N-1)$ 'relevant' eigenvalues of the rescaled adjacency matrix $A'_{ij} = A_{ij}/k_i$. One eigenvalue of this matrix is one, since $\sum_j A'_{ij} \equiv 1$, and we regard this eigenvalue as irrelevant, since it does not figure in the stability analysis[26]. If the $(N-1)$ relevant eigenvalues all fall within a range $(\lambda_-, \lambda_+)$ predicted by the analysis, then the synchronous solution is linearly stable. Note that the range $(\lambda_-, \lambda_+)$ is determined solely by the node dynamics and is independent of the network, while the relevant eigenvalues are determined solely by the network. In Fig. 4b, we plot the two relevant eigenvalues $\lambda_1$ and $\lambda_2$ of $\mathbf{A}'$ as a function of $A_{12}$. The network only synchronizes when $\lambda_1$ and $\lambda_2$ are both within the specific interval $(\lambda_-, \lambda_+)$. The derivation of the master stability function used to obtain $\lambda_+$ and $\lambda_-$ for our network of optoelectronic oscillators with adaptive coupling strengths is presented in Appendix C.

The model described in Section II and Appendices B and C assumes a network of identical nodes. In any practical realization, parameter mismatches are unavoidable and will affect the synchronization behavior of the network. An important and non-trivial question is how this adaptive strategy theory applies when there are small deviations from identicality. Here, we find good agreement between our theoretical calculation which assumes identical nodes and experiments where there is some parameter tolerance. We conclude that the theory, as presented, is applicable for making realistic predictions about which actual networks will synchronize. Moreover, our stability theory, based on the master stability function formalism, can be extended to the case of mismatches in a network of nearly identical systems[29,30].

## VI. NUMERICAL EXPERIMENTS ON A 25 NODE NETWORK

As a further example, we numerically consider a network of optoelectronic systems in which we have a relatively large number of nodes ($N = 25$) and in which the couplings are directional; i.e. $A_{ij}$ may differ from $A_{ji}$. For cases with $\mathbf{A}$ not symmetric the eigenvalues of $\mathbf{A}'$ may be complex. The two contours in Fig. 5b indicate the region of the complex plane within which the eigenvalues of $\mathbf{A}'$ must fall in order to maintain synchrony, i.e., the region for which the master stability function, $M(\lambda)$ is negative. The outer contour was obtained with $z_0 = 0.99$ and the inner contour corresponds to $z_0 = 0.95$. These correspond to smoothing times of 4.17 ms and 0.83 ms, respectively. Fig. 5a is a cut along the real axis and values of $M$ are plotted in $\text{ms}^{-1}$. The upper and lower critical bounds are labeled $\lambda_+$ and $\lambda_-$ and correspond to the horizontal lines in Fig. 4b. When constructing the adjacency matrix, we randomly choose each of the $A_{ij}$ to be either 1/7 or 0, with probability 0.25 and 0.75, respectively; the diagonal elements $A_{ii}$ are 1/7. This choice of adjacency matrix yields a distribution of row sums for which the adaptive algorithm must compensate. A graph of this asymmetric network is depicted in Fig. 6. The relevant eigenvalues of $\mathbf{A}'$ are plotted in the complex plane in Fig. 5b. The $(N-1)$ relevant eigenvalues fall within the region for stable synchrony when $z_0 = 0.99$ but not when $z_0 = 0.95$. In Fig. 7a, we initially run the full network of $N$ nodes without enabling the adaptive strategy. For $t < 0$, the synchronization error is large. The control is enabled at $t = 0$. The adaptive algorithm pulls the network into synchrony from initially uncorrelated states with an exponential convergence rate, related to the eigenvalue of $\mathbf{A}'$ with the largest associated relevant Lyapunov exponent. The same numerical experiment was performed with $z_0 = 0.95$ shown in Fig. 7b. In this case, failure to achieve synchronization was observed as predicted by the master stability function.

## VII. CONCLUSION

In summary, we have demonstrated an adaptive strategy for maintaining synchronization in a network of chaotic oscillators which we have analyzed using the master stability function technique extended to incorporate our adaptive strategy. This scheme is implemented in a network of three bidirectionally coupled chaotic optoelectronic feedback loops. Our experiments show that the adaptive strategy provides the ability to sense and track variations of the network couplings.

## ACKNOWLEDGMENTS

We thank Caitlin Williams and Karl Schmitt for advice and help. This work is supported by DOD MURI grant (ONR N000140710734).

## APPENDIX A: DISCRETE TIME EQUATIONS FOR AN ISOLATED OPTOELECTRONIC OSCILLATOR

The dynamical behavior of a nonlinear optoelectronic feedback loop like those considered here can be described by a pair of coupled nonlinear delay differential equations[15,20–23],

$$\begin{aligned} \tau_L \frac{du(t)}{dt} &= -\Big(1 - \frac{\tau_L}{\tau_H}\Big) u(t) - v(t) + \gamma \cos^2[u(t-\tau) + \phi], \\ \tau_H \frac{dv(t)}{dt} &= u(t), \end{aligned} \tag{7}$$

where $u(t)$ is a normalized output signal representing the voltage applied to the modulator, $\gamma$ is a dimensionless feedback constant, $\tau$ is the net time delay of the feedback path, and $\phi$ is the bias point of the Mach-Zehnder modulator. $\tau_L$ and $\tau_H$ are time constants for cascaded single-pole low-pass and high-pass filters. Eqs. (7) represent a band-pass filter with output $u(t)$ and input given by the self-feedback term $w(t) = \gamma \cos^2[u(t-\tau)+\phi]$. The linear filter operation can also be expressed in the Laplace domain,

$$U(s) = H_A(s)W(s)$$

where $H_A(s)$ is the continuous time transfer function of the filter, given by

$$H_A(s) = \frac{s\tau_H}{(1+s\tau_L)(1+s\tau_H)}. \tag{8}$$

It is noted that in order to fully describe the state of the isolated system at time $t$, one must have knowledge of $u(t)$ and history of $w(t)$ over the continuous interval $[t-\tau, t]$, making the system infinite dimensional in principle[24].

In this Appendix, we outline the mathematical structure for transforming Eqs. (7) into discrete time, as implemented in this set of experiments employing digital signal processing (DSP) technology. Our digital filter was designed to act as a two-pole band-pass filter ($M = 2$) that approximates the continuous time filter described by Eq. (8). The discrete time transfer function $H_D(z)$ is obtained from the continuous time function $H_A(s)$ by applying a bilinear transform with frequency pre-warping[27]. This process yields the equivalent discrete time transfer function,

$$H_D(z) = \frac{1}{4}(1-z_L)(1+z_H)\frac{(1-z^{-2})}{(1+z_L z^{-1})(1+z_H z^{-1})}, \tag{9}$$

where $z_L$ and $z_H$ are the poles of the discrete time filter, which are related to the time constants $\tau_L$ and $\tau_H$ and the sampling periods $T_s = 1/f_s$ by

$$z_H = \frac{1-\tan(T_s/2\tau_H)}{1+\tan(T_s/2\tau_H)}, \; z_L = \frac{1-\tan(T_s/2\tau_L)}{1+\tan(T_s/2\tau_L)}.$$

The discrete time filter can be represented in the time-domain by the linear difference equation,

$$u(n) = -a_1 u(n-1) - a_2 u(n-2) + b_0 w(n) + b_1 w(n-1) + b_2 w(n-2), \tag{10}$$

where the filter parameters are given by

$$a_1 = -(z_H + z_L), a_2 = z_H z_L,$$
$$b_0 = \tfrac{1}{4}(1-z_H)(1-z_L), b_1 = 0, b_2 = -b_0,$$

and the filter input is $w(n) = g(u(n-d)) \equiv \gamma \cos^2[u(n-d)+\phi]$ and $d = \tau/T_s$ is an integer describing the time delay, which we take to be an integer number of time-steps. These coefficients can be easily obtained for many choices of filter using MATLAB's Filter Design and Analysis Tool[28] (FDATool).

Tables I and II list all the parameters used in the experiments and simulations for continuous time and discrete time representations.

## APPENDIX B: ADAPTIVE SYNCHRONIZATION STRATEGY FOR A NETWORK OF OPTOELECTRONIC OSCILLATORS

In this Appendix, we describe a coupled network of $N$ identical optoelectronic nodes which implement the adaptive synchronization control rule. Next, we cast our system into the general form for discrete time coupled oscillators as given in Section II.

In our experiments, each of the three coupled nodes receives a superposition of optical signals from the two other nodes: $\sum_{j \neq i} A_{ij}(n) \cos^2[u_j(n-d)+\phi]$. In our implementation, a DSP is employed at each node to process this received signal and incorporate it into the local dynamics. The two input channels of the analog-to-digital converter are AC coupled, so we do not have direct access to this received optical signal or the locally generated optical signal, $\cos^2[u_i(n)+\phi]$. Our DSP boards are programed to perform a band-pass filtering routine on each signal and sum the outputs. For slowly-varying $A_{ij}(n)$, the linear filtering and weighted summation operations can be commuted. The cumulative input signal is well approximated as $r_i(n) = \sum_{j=1}^{N} A_{ij} u_j(n)$. We note that all the analytic and numerical results presented here apply under the assumption of slowly-varying or static coupling strengths.

Under this assumption, the equations for $N$ coupled optoelectronic nodes are:

Internal feedback

$$u_i(n) = -a_1 u_i(n-1) - a_2 u_i(n-2) + b_0 g(y_i(n-d)) - b_0 g(y_i(n-d-2)), \tag{11a}$$

Coupling term

$$r_i(n) = \sum_{j=1}^{N} A_{ij} u_j(n) \tag{11b}$$

Adaptive strategy

$$p_i(n) = z_0 p_i(n-1) + (1-z_0) r_i(n) u_i(n), \tag{11c}$$

$$q_i(n) = z_0 q_i(n-1) + (1-z_0) r_i^2(n), \tag{11d}$$

Weighted coupling term

$$y_i(n) = \frac{p_i(n)}{q_i(n)} r_i(n), \tag{11e}$$

We can convert these delay difference equations (Eqs. (11)) into a nonlinear map by constructing a state vector $\mathbf{x}_i(n)$ that is comprised of the band-pass filter output terms $u_i(n-1)$ and $u_i(n-2)$ and a history of the $(d+M)$ most recent elements of the filter input $y_i$:

$$\mathbf{x}_i(n) = \Big[ u_i(n-1) \;\; u_i(n-2) \;\; y_i(n-1) \;\; y_i(n-2) \;\; \cdots \;\; y_i(n-d-M) \Big]^T \tag{12}$$

(where $M = 2$ is the order of the filter). With this definition, the map equations for the coupled $\mathbf{x}_i$'s are expressed as

$$\mathbf{x}_i(n+1) = \begin{bmatrix} H(\mathbf{x}_i(n)) \\ u_i(n-1) \\ 0 \\ y_i(n-1) \\ y_i(n-2) \\ \vdots \\ y_i(n-d-1) \end{bmatrix} + \begin{bmatrix} 0 \\ 0 \\ 1 \\ 0 \\ 0 \\ \vdots \\ 0 \end{bmatrix} \frac{p_i(n)}{q_i(n)} \sum_{j=1}^{N} A_{ij} H(\mathbf{x}_j(n)), \tag{13}$$

where the function $H(\mathbf{x}_i(n))$ represents a scalar 'observable' component of the signal,

$$H(\mathbf{x}_i(n)) = -a_1 u_i(n-1) - a_2 u_i(n-2) + b_0 [g(y_i(n-d)) - g(y_i(n-d-2))]. \tag{14}$$

For an isolated node, $H(\mathbf{x}(n))$ is simply $u(n)$, the output of the discrete time filter. This map equation (13) is of the form of Eq. (2), by setting $\alpha_0 = 1$.

## APPENDIX C: STABILITY ANALYSIS OF SYNCHRONOUS SOLUTION

In this Appendix, we derive the master stability function for $N$ coupled systems described by Eq. (2) under the control of the adaptive rule given by Eq. (5). We follow the approach of Ref. 12, but here described in discrete time.

Eqs. (2) admit the synchronous solution $\mathbf{x}_i(n) = \mathbf{x}_s(n)$, for all $i$ and $n$, given by.

$$\begin{aligned} \mathbf{x}_s(n+1) &= \mathbf{F}(\mathbf{x}_s(n)) + \mathbf{v}\alpha_0 H(\mathbf{x}_s(n)), \\ p_i(n) &= k_i \langle H(\mathbf{x}_s(n))^2 \rangle_{z_0}, \\ q_i(n) &= k_i^2 \langle H(\mathbf{x}_s(n))^2 \rangle_{z_0}, \end{aligned} \tag{15}$$

where $k_i = \sum_{j=1}^N A_{ij}$. Note that in the synchronous state, $q_i/p_i = \overline{k}_i = k_i$. By linearizing (2) about the synchronous solution (15), we obtain,

$$\begin{aligned} \delta\mathbf{x}_i(n+1) &= \mathbf{DF}(\mathbf{x}_s(n))\delta\mathbf{x}_i(n) + \mathbf{v}\alpha_0 \Big[ \frac{H(\mathbf{x}_s(n))}{k_i^2 \langle (H(\mathbf{x}_s(n))^2 \rangle_{z_0}} \epsilon_i(n) + \frac{\mathbf{D}H(\mathbf{x}_s(n))}{k_i} \sum_{j=1}^N A_{ij}\delta\mathbf{x}_j(n) \Big], \\ \epsilon_i(n+1) &= z_0\epsilon_i(n) + (1-z_0)k_i \Big[ k_i\delta\mathbf{x}_i(n) - \sum_{j=1}^N A_{ij}\delta\mathbf{x}_j(n) \Big] H(\mathbf{x}_s(n))\mathbf{D}H(\mathbf{x}_s(n)), \end{aligned} \tag{16}$$

$i = 1, \dots, N$, where we have introduced the variable $\epsilon_i \equiv k_i \delta p_i - \delta q_i$ and $\mathbf{DF} \equiv \partial\mathbf{F}/\partial\mathbf{x}$, $\mathbf{D}H \equiv \partial H/\partial\mathbf{x}$. Eqs. (16) constitute a system of $mN$ coupled equations where $\mathbf{x}_i$ is a state vector of length $m$. In order to simplify the analysis, we seek to decouple this system into $N$ independent systems, each of dimension $m$. For this purpose we seek a solution where $\delta\mathbf{x}_i(n)$ is in the form $\delta\mathbf{x}_i(n) = c_i\delta\overline{\mathbf{x}}(n)$, where $c_i$ is a time independent scalar that depends on $i$ and $\delta\overline{\mathbf{x}}(n)$ depends on time but not on $i$. Substituting this ansatz into Eqs. (16), we obtain,

$$\begin{aligned} \delta\overline{\mathbf{x}}(n+1) &= \mathbf{DF}(\mathbf{x}_s(n))\delta\overline{\mathbf{x}}(n) + \mathbf{v}\frac{\alpha_0}{c_i} \Big[ \frac{H(\mathbf{x}_s(n))}{k_i^2 \langle (H(\mathbf{x}_s(n))^2 \rangle_{z_0}} \epsilon_i(n) + \frac{\mathbf{D}H(\mathbf{x}_s(n))}{k_i} \delta\overline{\mathbf{x}}(n) \sum_{j=1}^N A_{ij}c_j \Big], \\ \epsilon_i(n+1) &= z_0\epsilon_i(n) + (1-z_0)k_i \Big[ k_i c_i - \sum_{j=1}^N A_{ij}c_j \Big] H(\mathbf{x}_s(n))\mathbf{D}H(\mathbf{x}_s(n))\delta\overline{\mathbf{x}}(n), \end{aligned} \tag{17}$$

To make Eqs. (17) independent of $i$, we consider $\overline{\epsilon}(n) = \epsilon_i(n)/[(1-\lambda)k_i^2 c_i]$ and $\Sigma_j A_{ij}c_j = \lambda k_i c_i$, where $\lambda$ is a quantity independent of $i$. Namely, the possible values of $\lambda$ are the eigenvalues $\mathbf{A}'\mathbf{c} = \lambda\mathbf{c}$, corresponding to linearly independent eigenvectors $\mathbf{c} = \Big[ c_1 \ c_2 \ \cdots \ c_N \Big]$, where $\mathbf{A}' = \{A'_{ij}\} = \{k_i^{-1} A_{ij}\}$. This gives,

$$\begin{aligned}
\delta\overline{\mathbf{x}}(n+1) &= \mathbf{DF}(\mathbf{x}_s(n))\delta\overline{\mathbf{x}}(n) + \mathbf{v}\alpha_0\Big[(1-\lambda)\frac{H(\mathbf{x}_s(n))}{\langle (H(\mathbf{x}_s(n))^2\rangle_{z_0}}\overline{\epsilon}(n) + \lambda \mathbf{D}H(\mathbf{x}_s(n))\delta\overline{\mathbf{x}}(n)\Big], \\
\overline{\epsilon}(n+1) &= z_0\overline{\epsilon}(n) + (1-z_0)H(\mathbf{x}_s(n))\mathbf{D}H(\mathbf{x}_s(n))\delta\overline{\mathbf{x}}(n),
\end{aligned} \tag{18}$$

which is independent of $i$ but depends on the eigenvalues $\lambda$. Considering the typical case where there are $N$ distinct eigenvalues of the $N \times N$ matrix $\mathbf{A}'$, we see that Eqs. (18) constitute $N$ decoupled linear difference equations for the synchronization perturbation variables $\delta\overline{\mathbf{x}}$ and $\overline{\epsilon}$. All the rows of $\mathbf{A}'$ sum to 1, therefore $\mathbf{A}'$ has at least one eigenvalues $\lambda = 1$, corresponding to the eigenvector $c_1 = c_2 = \ldots = c_N = 1$. For $\lambda = 1$, Eqs. (18) becomes

$$\delta\overline{\mathbf{x}}(n+1) = \Big[\mathbf{DF}(\mathbf{x}_s(n)) + \mathbf{v}\alpha_0\mathbf{D}H(\mathbf{x}_s(n))\Big]\delta\overline{\mathbf{x}}(n). \tag{19}$$

This equation reflects the chaos of the reference synchronized state (15) and (because all the $c_i$ are equal) is associated with perturbations which are tangent to the synchronization manifold and are therefore irrelevant in determining synchronization stability. Stability of the synchronized state thus requires Eq. (18) yield exponential decay of $\delta\overline{\mathbf{x}}$ and $\overline{\epsilon}$ for all the eigenvalues $\lambda$, excluding the 'irrelevant' $\lambda = 1$ eigenvalue.

The problem of stability of the synchronized solutions (15) for an arbitrary network of coupled systems evolving according to Eq. (2) can be evaluated using a master stability function $M(\lambda, z_0)$ that associates the pair $(\lambda, z_0)$ with the maximum Lyapunov exponent of Eqs. (18).

For a network of optoelectronic feedback loops described by the Eqs. (13) and (14), we define the variational vector as:

$$\delta\overline{\mathbf{x}}(n) = \Big[\delta\overline{u}(n-1)\ \ \delta\overline{u}(n-2)\ \ \delta\overline{y}(n-1)\ \ \delta\overline{y}(n-2)\ \ \cdots\ \ \delta\overline{y}(n-d-M)\Big]^T. \tag{20}$$

The variational equations describing perturbations from the synchronous solution (Eqs. (18) with the specific $\mathbf{F}$, $H$, $\mathbf{v}$, and $\alpha_0$ for our experimental system) are:

$$\begin{aligned}
\overline{\epsilon}(n) &= z_0\overline{\epsilon}(n-1) + (1-z_0)u_s(n-1)\delta\overline{u}(n-1), \\
\delta\overline{u}(n) &= -a_1\delta\overline{u}(n-1) - a_2\delta\overline{u}(n-2) + b_0\Big[\delta g(u_s(n-d))\delta\overline{y}(n-d) - \delta g(u_s(n-d-2))\delta\overline{y}(n-d-2)\Big], \\
\delta\overline{y}(n) &= (1-\lambda)\frac{u_s(n)}{\langle u_s(n)^2\rangle_{z_0}}\overline{\epsilon}(n) + \lambda\delta\overline{u}(n), \\
\delta g(u) &\equiv \frac{\partial g}{\partial u} = -2\gamma\sin(u+\phi)\cos(u+\phi).
\end{aligned} \tag{21}$$

These are conditional upon the synchronous solution:

$$u_s(n) = -a_1 u_s(n-1) - a_2 u_s(n-2) + b_0 g(u_s(n-d)) - b_0 g(u_s(n-d-2)). \qquad (22)$$

The master stability function $M(\lambda, z_0)$ for $z_0 = 0.95$ and $z_0 = 0.99$, obtained by numerically iterating Eqs. (21) and (22), is presented in Fig. 5.

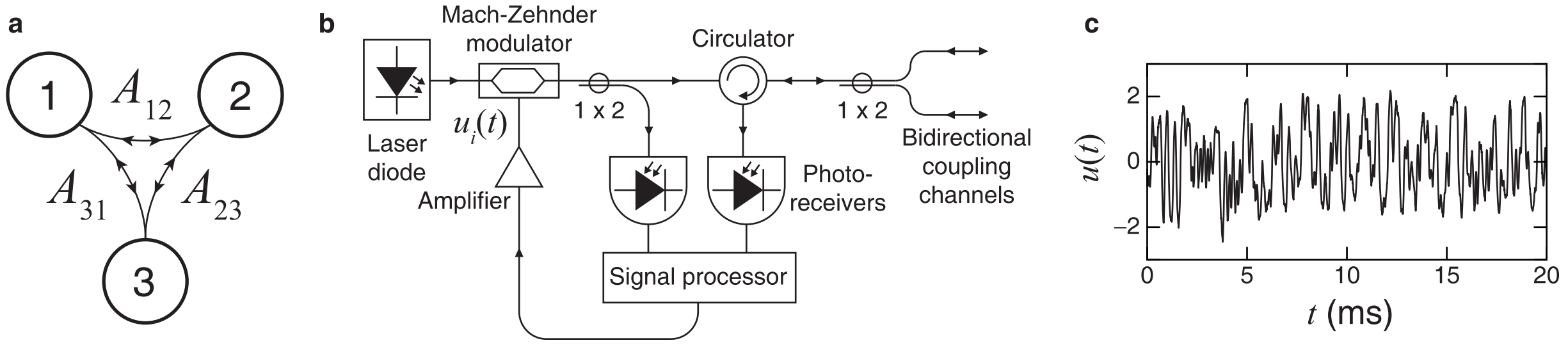

FIG. 1. (a) Topology of bidirectionally coupled three-node network where symmetric coupling strengths $A_{ij}$ are controlled by three independent variable optical attenuators. (b) Experimental schematic of a single optoelectronic node. (c) A representative time series of an isolated feedback loop.

| Parameter | Value | Unit |
|---|---|---|
| $f_H = (2\pi\tau_H)^{-1}$ | 100 | Hz |
| $f_L = (2\pi\tau_L)^{-1}$ | 2.5 | kHz |
| $\tau$ | 1.5 | ms |
| $\phi$ | $-\pi/4$ | rad. |
| $\gamma$ | 4.5 | - |

TABLE I. System parameters in continuous time representation

| Parameter | Value | Unit |
|---|---|---|
| $f_s = 1/T_s$ | 24 | kSamples/s |
| $d$ | 36 | Samples |
| $a_1$ | –1.4962 | - |
| $a_2$ | 0.5095 | - |
| $b_0 = -b_2$ | 0.2452 | - |

TABLE II. System parameters in discrete time representation

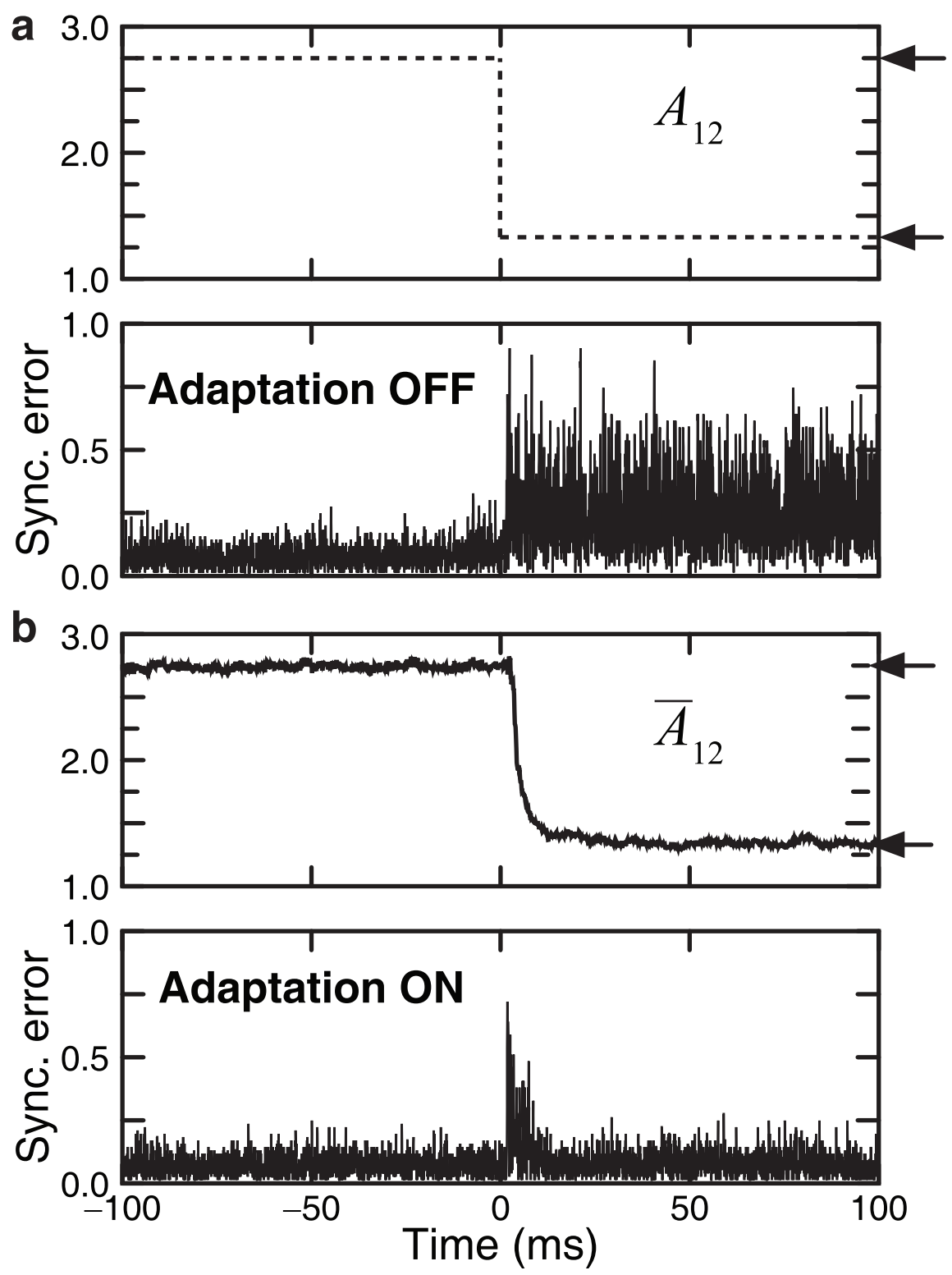

FIG. 2. (a) The uncontrolled network is initially tuned to synchrony. At $t = 0$, the coupling strength $A_{12}$ is suddenly decreased. Synchrony is broken between all three nodes, shown by a large synchronization error for $t > 0$. (b) The same situation is explored as in (a), but the adaptive synchronization algorithm is enabled at each node. Now, when the coupling strength is changed, we are able to obtain an estimate (denoted $\overline{A}_{12}$) of $A_{12}$ and to maintain synchrony in the network.

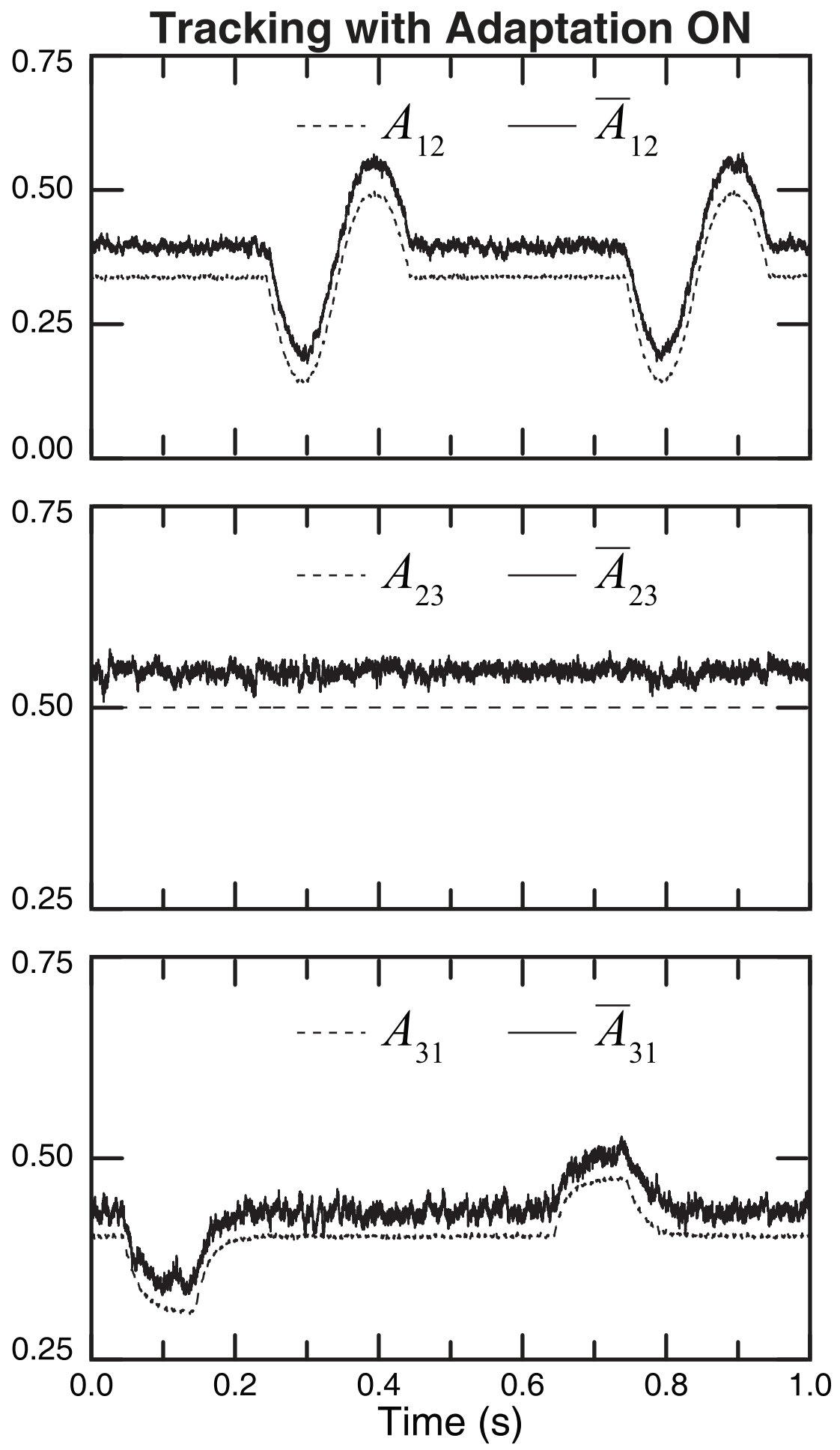

FIG. 3. An experiment where $A_{12}(t)$ and $A_{31}(t)$ are smoothly varying in time. The algorithm tracks how each element of the coupling matrix is varying. The coupling strength variations along with the respective estimates (artificially displaced by 0.05 to distinguish the traces) are shown for $z_0 = 0.99$ corresponding to a smoothing time of 4.17 ms.

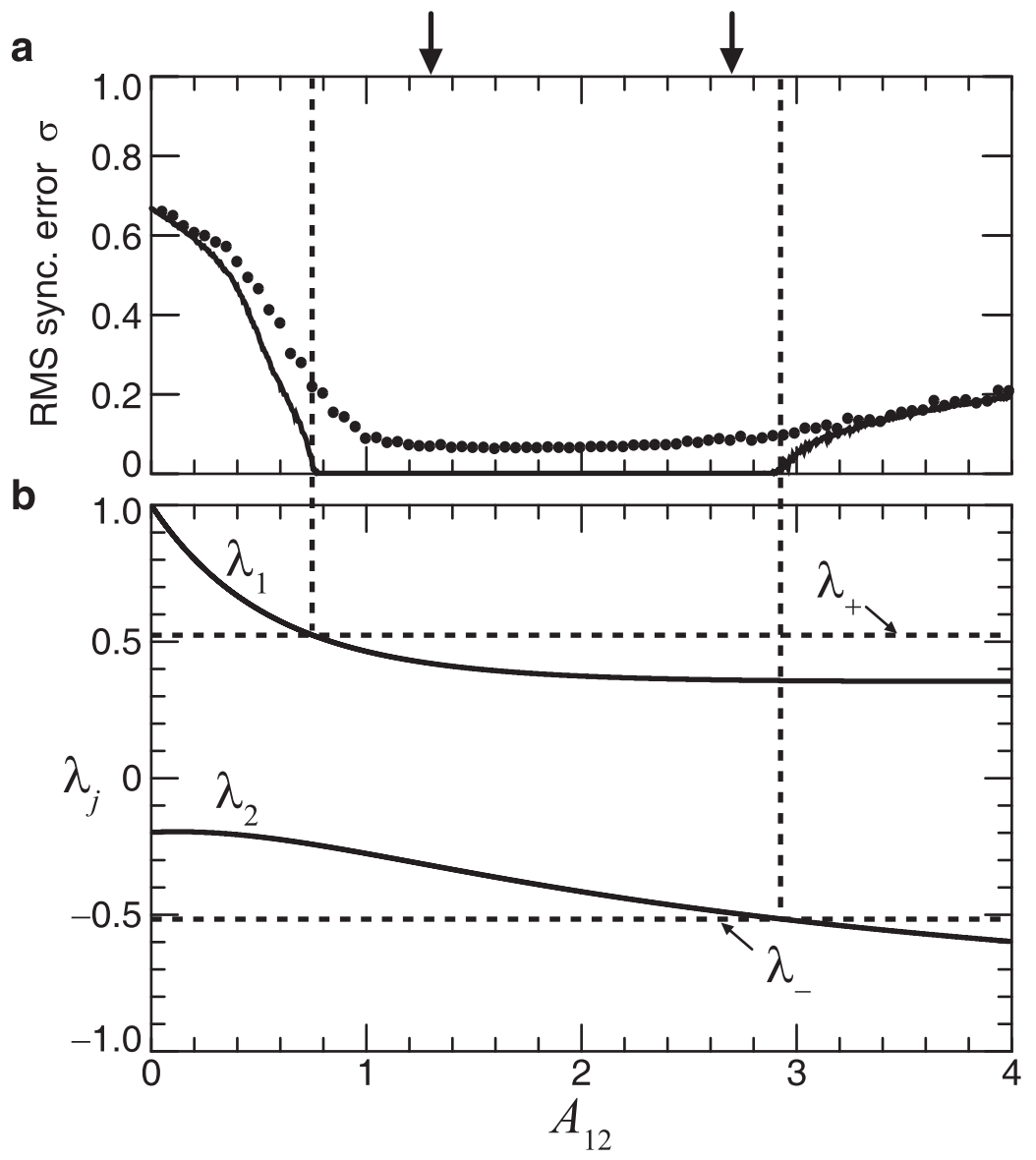

FIG. 4. (a) Measurement of the global RMS synchronization error $\sigma$ versus $A_{12}$ with $A_{31} = 0$, $A_{23} = 1.5$ from experiments (points) and simulations (solid lines). The upper and lower bounds on $A_{12}$ for network synchronization given by the master stability function are shown by dashed vertical lines. The arrows above the plot point to the two values for $A_{12}$ where synchrony was observed in Fig. 2b (in Fig. 2b these values are indicated by horizontal arrows). (b) The calculated eigenvalues of the weighted adjacency matrix are plotted as we vary $A_{12}$. Both eigenvalues must fall within the range $(\lambda_-, \lambda_+)$ for stability of global synchronization.

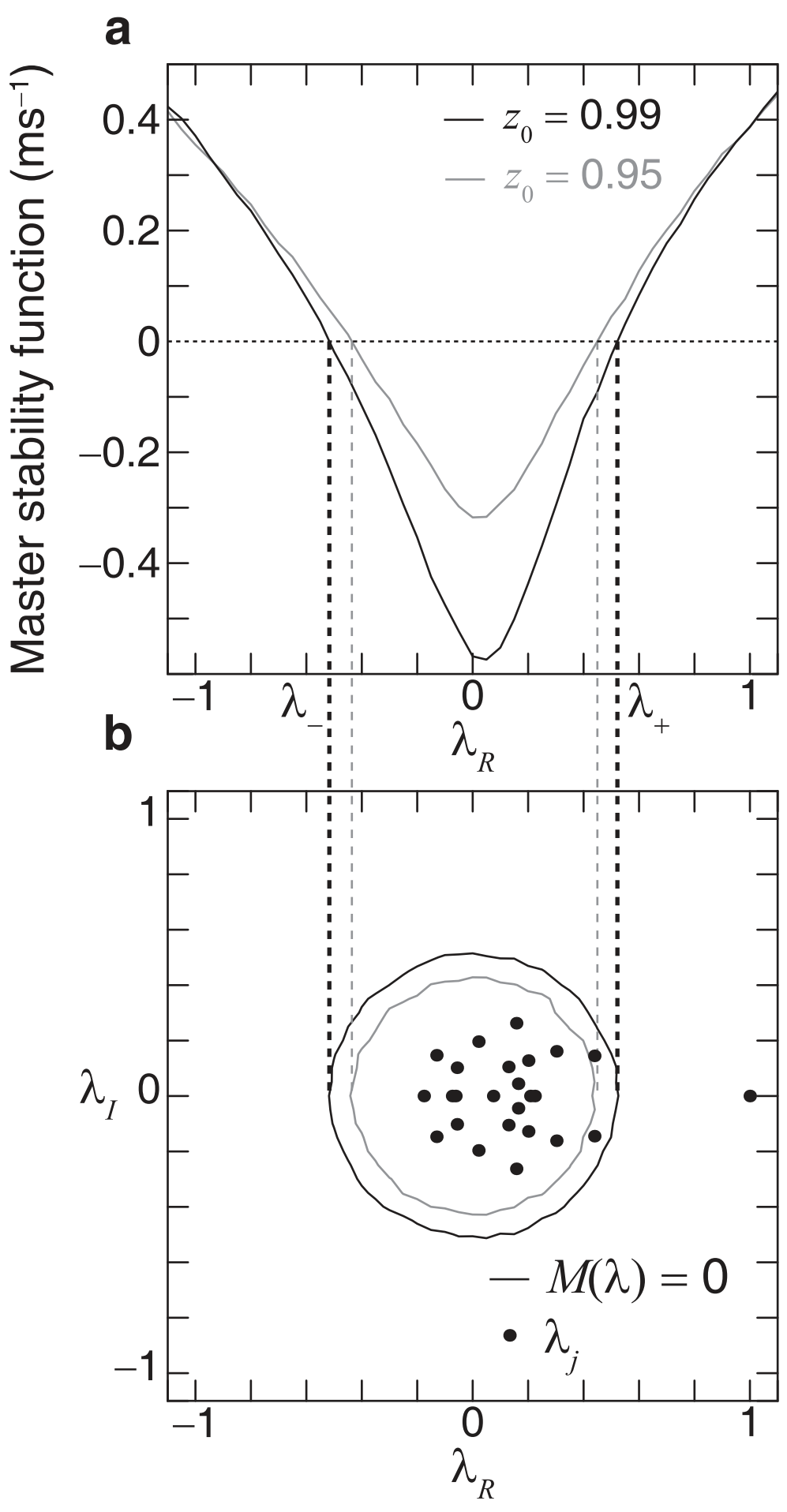

FIG. 5. (a) Master stability functions $M(\lambda)$ for real eigenvalues for $z_0 = 0.95$ and $z_0 = 0.99$. $\lambda_+$ and $\lambda_-$ are the critical values at which $M = 0$. (b) The solid curves are the $M = 0$ contour plots in the complex plane for these two values of $z_0$. Relevant eigenvalues, $\lambda_R + i\lambda_I$ , of $\mathbf{A}'$ must fall within the areas bounded by these curves for stable synchronization. The points are the 24 relevant eigenvalues for the $N = 25$ node network depicted in Fig. 6, which fall within the stable region for $z_0 = 0.99$ but not for $z_0 = 0.95$.

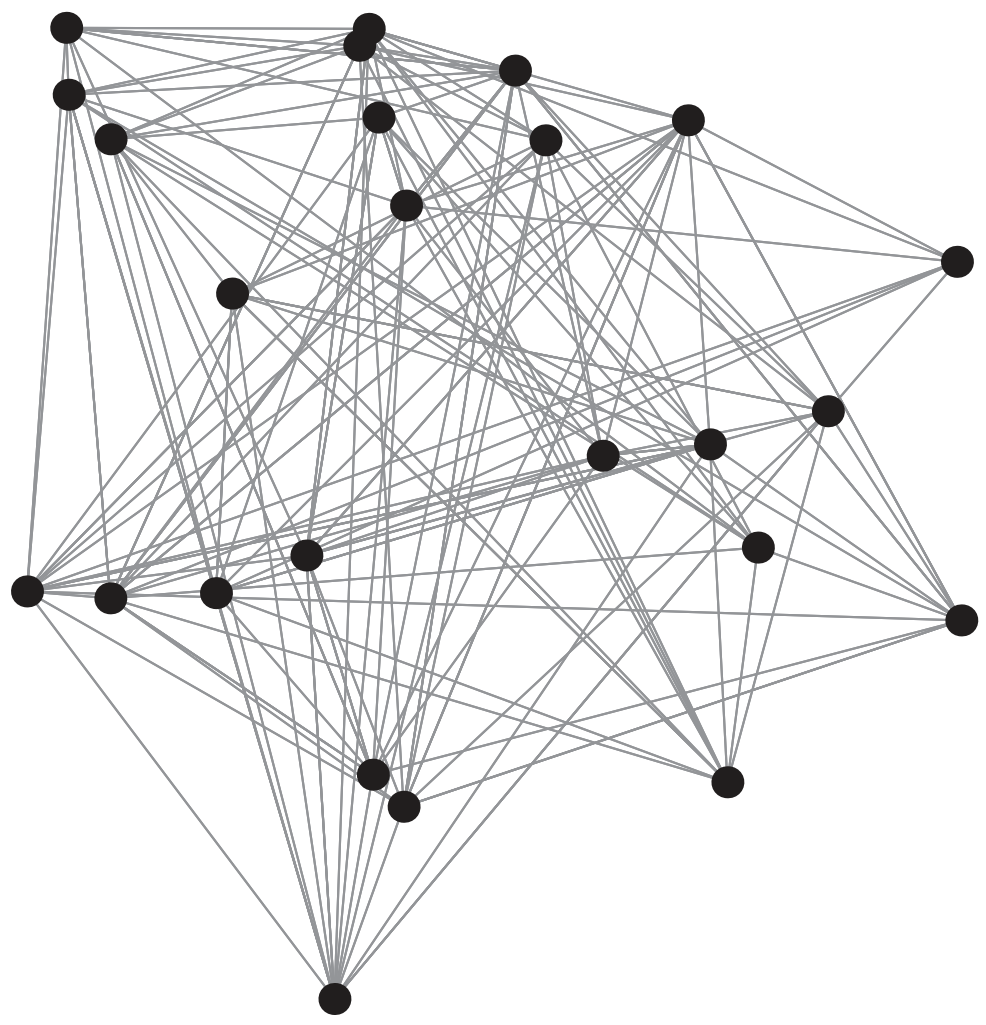

FIG. 6. A schematic representation of the random network of $N = 25$ coupled nodes used in the numerical experiment described in Sec. VI.

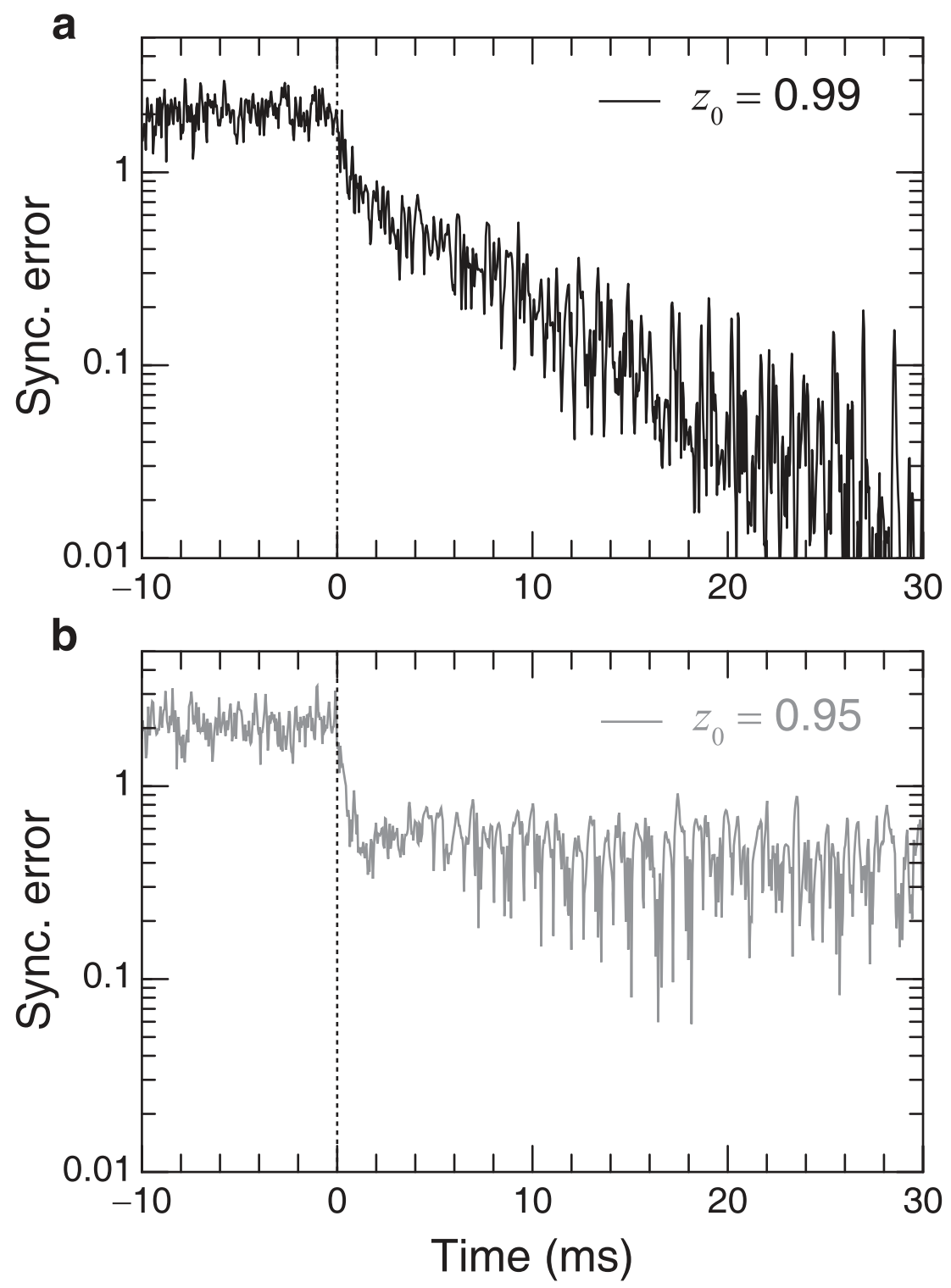

FIG. 7. Synchronization error in a numerical simulation of the 25 node network with (a) $z_0 = 0.99$ and (b) $z_0 = 0.95$. The adaptive algorithm is enabled at $t = 0$ and the global synchronization error decreases exponentially at a rate given by the master stability function for case (a) but not case (b).